\documentclass[10pt,conference, anonymous, review]{IEEEtran}
\IEEEoverridecommandlockouts

\newcommand{\code}[1]{\texttt{\small#1}} 

\usepackage{enumitem}
\setlist{leftmargin=3.7mm}
\usepackage{booktabs}
\usepackage{url}
\usepackage{amsthm}
\theoremstyle{definition}
\newtheorem{definition}{Definition}
\usepackage{amsmath}
\usepackage{booktabs}
\usepackage{graphicx}
\usepackage{tikz}
 
\usepackage{multirow}
\usepackage{listings}
\usepackage{tcolorbox}
\usepackage{color}
\usepackage{xcolor}
\usepackage{caption}
\usepackage{subcaption}
\usepackage{geometry}
\usepackage{longtable}
 \usepackage[normalem]{ulem}
 \useunder{\uline}{\ul}{}

\definecolor{deepblue}{rgb}{0,0,0.5}
\definecolor{deepgreen}{rgb}{0,0.5,0}
\definecolor{deepred}{rgb}{0.6,0,0}
\definecolor{darkorange}{RGB}{255,140,0}
\definecolor{lightgray}{rgb}{0.93,0.93,0.93}
\definecolor{deepgray}{rgb}{0.25,0.25,0.25}

\lstdefinelanguage{JSON}{
	keywords={thoughts, command, name, args},
	ndkeywords={},
	ndkeywordstyle=\color{deepgreen}\bfseries,
	basicstyle=\footnotesize\ttfamily,
	numberstyle=\color{deepgray},
	stepnumber=1,
	numbersep=8pt,
	showstringspaces=false,
	breaklines=true,
	frame=lines,
	backgroundcolor=\color{lightgray},
	commentstyle=\color{deepgreen},
	keywordstyle=\color{deepblue},
	stringstyle=\color{deepgreen},
	tabsize=4,
	captionpos=b,
	morecomment=[s]{/*}{*/},
	morestring=[b]",
	morestring=[d]',
	literate=
	*{[}{{{\color{deepblue}{\textbf{[}}}}}{1}
	{]}{{{\color{deepblue}{\textbf{]}}}}}{1}
	{:}{{{\color{deepblue}{\textbf{:}}}}}{1}
	{,}{{{\color{deepgreen}{,}}}}{1}
	{\{}{{{\color{deepblue}{\textbf{\{}}}}}{1}
	{\}}{{{\color{deepblue}{\textbf{\}}}}}}{1}
	{"*}{{{\color{deepgreen}{"*}}}}{1},
}

\lstdefinelanguage{Java}{
	basicstyle=\small\ttfamily,
	numberstyle=\color{deepgray},
	stepnumber=1,
	numbersep=8pt,
	showstringspaces=false,
	breaklines=true,
	frame=lines,
	backgroundcolor=\color{lightgray},
	commentstyle=\color{deepgreen},
	keywordstyle=\color{deepblue},
	stringstyle=\color{deepred},
	tabsize=4,
	captionpos=b,
	morekeywords={public, class, void, int, if, else, for, while, return, true, false},
	emph={String, System},
	emphstyle=\color{darkorange},
	alsoletter={.,;:[]()},
}

\newcommand{\name}{Repair\-Agent}

\begin{document}

\title{\name{}: An Autonomous, LLM-Based Agent for Program Repair}

\author{
Islem Bouzenia\\
University of Stuttgart\\
Germany\\
fi\textunderscore bouzenia@esi.dz\\

\and
Premkumar Devanbu\\
UC Davis\\
USA\\
ptdevanbu@ucdavis.edu\\

\and
Michael Pradel\\
University of Stuttgart\\
Germany\\
michael@binaervarianz.de\\
}

\maketitle

\begin{abstract}
Automated program repair has emerged as a powerful technique to mitigate the impact of software bugs on system reliability and user experience.
This paper introduces \name{}, the first work to address the program repair challenge through an autonomous agent based on a large language model (LLM).
Unlike existing deep learning-based approaches, which prompt a model with a fixed prompt or in a fixed feedback loop, our work treats the LLM as an agent capable of autonomously planning and executing actions to fix bugs by invoking suitable tools.
\name{} freely interleaves gathering information about the bug, gathering repair ingredients, and validating fixes, while deciding which tools to invoke based on the gathered information and feedback from previous fix attempts.
Key contributions that enable \name{} include a set of tools that are useful for program repair, a dynamically updated prompt format that allows the LLM to interact with these tools, and a finite state machine that guides the agent in invoking the tools.
Our evaluation on the popular Defects4J dataset demonstrates \name{}'s effectiveness in autonomously repairing 164 bugs, including 39 bugs not fixed by prior techniques.
Interacting with the LLM imposes an average cost of 270,000 tokens per bug, which, under the current pricing of OpenAI's GPT-3.5 model, translates to 14 cents per bug.
To the best of our knowledge, this work is the first to present an autonomous, LLM-based agent for program repair, paving the way for future agent-based techniques in software engineering.
\end{abstract}

\section{Introduction}
\label{sec:intro}

Software bugs lead to system failures, security vulnerabilities, and compromised user experience.
Fixing bugs is a critical task in software development, but if done manually, demands considerable time and effort.
Automated program repair (APR) promises to dramatically reduce this effort by addressing the critical need for effective and efficient bug resolution in an automated manner.
Researchers and practitioners have explored various approaches to address the challenge of automatically fixing bugs~\cite{cacm2019-program-repair}, including techniques based on manually designed~\cite{LeGoues2012,Liu2019a} and (semi-)automatically extracted~\cite{Kim2013,oopsla2019,Bavishi2019} fix patterns, based on symbolic constraints~\cite{Nguyen2013b,xuan2016nopol,mechtaev2016angelix}, and various machine learning-based approaches~\cite{Long2016,Gupta2017,Tufano2019,Lutellier2020,Chen2021d,Li2020a,Zhu2021}.

The current state-of-the-art in APR predominantly revolves around large language models (LLMs).
The first generation of LLM-based repair uses a one-time interaction with the model, where the model receives a prompt containing the buggy code and produces a fixed version~\cite{10172803,Jiang2023}.
The second and current generation of LLM-based repair introduces iterative approaches, which query the LLM repeatedly based on feedback obtained from previous fix attempts~\cite{Xia2023a,Kang2023a,Ye2024}.

A key limitation of current iterative, LLM-based repair techniques is that their hard-coded feedback loops do not allow the model to gather information about the bug or existing code that may provide ingredients to fix the bug.
Instead, these approaches fix the code context that is provided in the prompt, typically to the buggy code~\cite{Xia2023a,Ye2024}, and sometimes also details about the test cases that fail~\cite{Kang2023a}.
The feedback loop then executes the tests on different variants of the buggy code and adds any compilation errors, test failures, or other output, to the prompt of the next iteration.
However, this approach fundamentally differs from the way human developers fix bugs, which typically involves a temporal interleaving of gathering information to understand the bug, searching code that may be helpful for fixing the bug, and experimenting with candidate fixes~\cite{ko2006exploratory,bohme2017bug}.

This paper presents \name{}, the first autonomous, LLM-based agent for automated program repair.
Our approach treats the LLM as an autonomous agent capable of planning and executing actions to achieve the goal of fixing a bug.
We equip the LLM with a set of bug repair-specific tools that the models can invoke to interact with the code base in a way similar to a human developer.
For example, \name{} has tools to extract information about the bug by reading specific lines of code, to gather repair ingredients by searching the code base, and to propose and validate fixes by applying a patch and executing test cases.
Importantly, we do not hard-code how and when to use these tools, but instead let the LLM autonomously decide which tool to invoke next, based on previously gathered information and feedback from previous fix attempts.

Our approach is enabled by three key components.
First, a general-purpose LLM, such as GPT-3.5, which we query repeatedly with a dynamically updated prompt.
We contribute a novel prompt format that guides the LLM through the bug repair process, and that gets updated based on the commands invoked by the LLM and the results of the previous command executions.
Second, a set of tools that the LLM can invoke to interact with the code base.
We present a set of 14 tools designed to cover different steps a human developer would take when fixing a bug, such as reading specific lines of code, searching the code base, and applying a patch.
Third, a middleware that orchestrates the communication between the LLM and the tools.
We present novel techniques for guiding tool invocations through a finite state machine and for heuristically interpreting possibly incorrect LLM outputs.
The iterative loop of \name{} continues until the agent declares to have found a suitable fix, or until exhausting a budget of iterations.

To evaluate the effectiveness of our approach, we apply it to all 835 bugs in the Defects4J~\cite{Just2014} dataset, a widely used benchmark for evaluating program repair techniques.
\name{} successfully fixes 164 bugs, including 74 and 90 bugs of Defects4J~v1.2 and~v2.0, respectively.
The correctly fixed bugs include 49 bugs that require fixing more than one line, showing that \name{} is capable of fixing complex bugs.
Compared to state-of-the-art techniques~\cite{Xia2023a,Ye2024}, \name{} successfully fixes 39 bugs not fixed by prior work.
Measuring the costs imposed by interacting with the LLM, we find that \name{} imposes an average cost of 270,000 tokens per bug, which, under the current pricing of OpenAI's GPT-3.5 model, translates to 14 cents per bug.
An additional evaluation on a set of recent bugs~\cite{gitbugjava} shows that \name{} is able to achieve similar performance on single-line bugs while being a bit worse on multi-line and multi-file bugs, mainly, due to a higher complexity of bugs in GitBug-Java dataset. We believe from these results that \name{} is not much affected by potential data leakage of Defects4J.
Overall, our results show that our agent-based approach establishes a new state of the art in program repair.

In summary, this paper contributes the following:
\begin{itemize}
	\item An autonomous, LLM-based agent for program repair.
	\item A dynamically updated prompt format that guides the LLM through the bug repair process.
	\item A set of tools that enable a LLM to to perform steps a human developer would take when fixing a bug.
	\item A middleware that orchestrates the communication between the LLM and the tools.
	\item Empirical evidence that \name{} establishes a new state of the art by successfully fixing 164 bugs, including 39 bugs not fixed by prior work.
	\item We will release the implementation of \name{} as open-source to foster future work.
\end{itemize}
To the best of our knowledge, there currently is no published work on an autonomous, LLM-based agent for any code-generation task.
We envision \name{} to pave the way for future agent-based techniques in software engineering.

\section{Background on LLM-Based, Autonomous Agents}

By virtue of being trained on vast amounts of web knowledge, including natural language and source code, LLMs have demonstrated remarkable abilities in achieving human-level performance for various tasks~\cite{Bubeck2023}.
A promising way of using these abilities are \emph{LLM-based agents}, by which we mean LLM-based techniques with two properties:
(1) The LLM autonomously plans and executes a sequence of actions to achieve a goal, as opposed to responding to a hard-coded query or being queried in a hard-coded algorithm.
(2) The actions performed by the LLM include invocations of external tools that enable the LLM to interact with its environment~\cite{Schick2023,Patil2023}.
In the context of software engineering, and automated repair in particular, the tools could be tools usually used by developers, e.g., as part of an integrated development environment (IDE).
The basic idea is to query the LLM with a prompt that contains the current state of the world, the goal to be achieved, and a set of actions that could be performed next.
The model than decides which action to perform, and the feedback from performing the action is integrated into the next prompt.
Recent surveys provide a comprehensive overview of LLM-based, autonomous agents~\cite{Wang2023} and of LLM agents equipped with tools invoked via APIs~\cite{Mialon2023}.
The potential of such agents for software engineering currently is not well explored, which this paper aims to address for the challenging task of automated program repair.

\section{Approach}

\subsection{Overview}
\label{sec: overview}

\begin{figure*}
	\centering
	\includegraphics[width=1\linewidth]{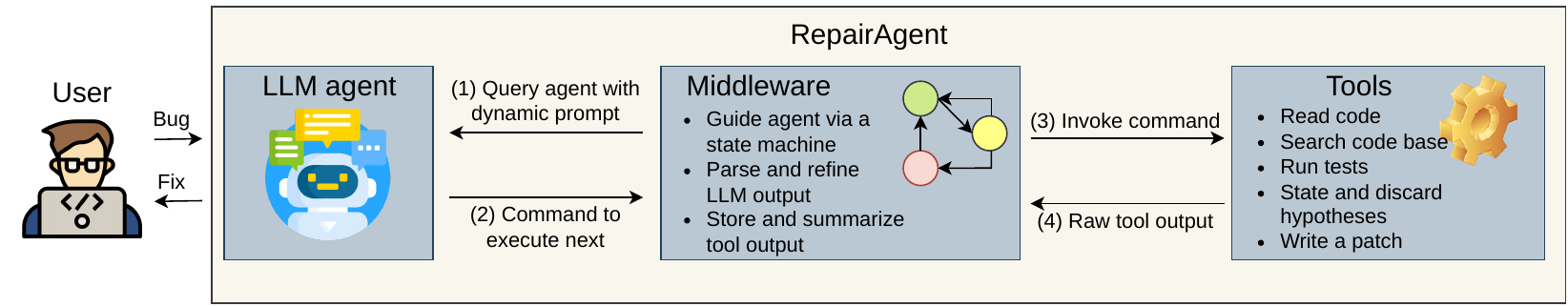}
	\caption[]{Overview of \name{}.}
	\label{fig:tooloverview}
\end{figure*}

Figure~\ref{fig:tooloverview} gives an overview of the \name{} approach, which consists of three components: an LLM agent (left), a set of tools (right), and a middleware that orchestrates the communication between the two (middle).
Given a bug to fix, the middleware initializes the LLM agent with a prompt that contains task information and instructions on how to perform it by using the provided tools (arrow~1).
The LLM responds by suggesting a call to one of the available tools (arrow~2), which the middleware parses and then executes (arrow~3).
The output of the tool (arrow~4) is then integrated into the prompt for the next invocation of the LLM, and the process continues iteratively until the bug is fixed or a predefined budget is exhausted.

\subsection{Terminology}

\name{} proceeds in multiple iterations, or cycles:

\begin{definition}[Cycle]
	\label{def:cycle}
	A \emph{cycle} represents one round of interaction with the LLM agent, which consists of the following steps:
	\begin{enumerate}
		\item Query the agent
		\item Post-process the response
		\item Execute the command suggested by the agent
		\item Update the dynamic prompt based on the command's output
	\end{enumerate}
\end{definition}

In each cycle, the approach queries the LLM once. 
The input to the model is updated based on commands (calls to tools) invoked by the LLM, and their results, in previous cycles.
We call the model input a dynamic prompt:

\begin{definition}[Dynamic prompt]
The \emph{dynamic prompt} is a sequence of text sections $P=[s_0, s_1,..., s_n]$, where each section $s_i$ is one of the following (where $s_i(c)$ refers to a section during a cycle $c$):
\begin{itemize}
	\item A \emph{static section}, which remains the same across all cycles, i.e., $s_i(c) = s_i(c')$ for all $c, c'$.
	\item A \emph{dynamic section}, which may differ across cycles, i.e., there may exist $c, c'$ with $s_i(c) \neq s_i(c')$.
\end{itemize}
\end{definition}

\subsection{Dynamic Prompting of the Repair Agent}

\begin{table}
	\caption{Sections of the dynamically updated prompt.}
	\label{tab:prompt}
	\centering
	\setlength{\tabcolsep}{25pt}
	\begin{tabular}{@{}ll@{}}
		\toprule
		Prompt section & Nature \\
		\midrule
		Role & Static \\
		Goals & Static \\
		Guidelines & Static \\
		State description & Dynamic \\
		Available tools & Dynamic \\
		Gathered information & Dynamic \\
		Specification of output format  & Static \\
		Last executed command and result  & Dynamic \\
		\bottomrule
	\end{tabular}
\end{table}

The repair agent is an LLM trained on natural language and source code, such as GPT-3.5.
\name{} queries the LLM with a dynamic prompt that consists of a sequence of static and dynamic sections, as listed in Table~\ref{tab:prompt}.
We describe each section in detail in the following.

\subsubsection{Role}

This section of the prompt defines the agent's area of expertise, which is to resolve bugs in Java code, and outlines the agent's primary objective: understanding and fixing bugs. 
The prompt emphasizes that the agent's decision-making process is autonomous and should not rely on user assistance.

\subsubsection{Goals}

We define five goals for the agent to pursue, which remain the same across all cycles:
\begin{itemize}
	\item \emph{Locate the bug:}
	Execute tests and use fault localization techniques to pinpoint the bug's location. Skip this goal when fault localization information is already provided in the prompt.
	
	\item \emph{Gather information about the bug:}
	Analyze the lines of code associated with the bug to understand the bug.
	
	\item \emph{Suggest simple fixes to the bug:}
	Start by suggesting simple fixes.
	
	\item \emph{Suggest complex fixes:}
	If simple fixes prove ineffective, explore and propose more complex ones.
	
	\item \emph{Iterate over the previous goals:}
	Continue to gather information and to suggest fixes until finding a fix.
\end{itemize}

\subsubsection{Guidelines}

We provide a set of guidelines.
First, we inform the model that there are diverse kinds of bugs, ranging from single-line issues to multi-line bugs that may entail changing, removing, or adding lines.
Based on the observation that many bugs can be fixed by relatively simple, recurring fix patterns~\cite{Karampatsis_Sutton_2020}, we provide a list of recurring fix patterns.
The list is based on the patterns described in prior work on single-statement bugs in Java~\cite{Karampatsis_Sutton_2020}.
For each pattern, we provide a short natural language description and an example of buggy and fixed code.
Second, we instruct the model to insert comments above the modified code, which serves two purposes.
On the one hand, the comments allow the model to explain its reasoning, which has been shown to enhance the reasoning abilities of LLMs~\cite{wei2022chain}.
On the other hand, commenting will ultimately help human developers in understanding the nature of the edits.
Third, we instruct the model to conclude its reasoning with a clearly defined next step that can be translated into a call to a tool.
Finally, we describe that there is a limited budget of tool invocations, highlighting the importance of efficiency in selecting the next steps.
Specifically, we specify a maximum number of cycles (40 by default).

\subsubsection{State Description}

\begin{figure}
	\centering
	\includegraphics[width=0.9\linewidth]{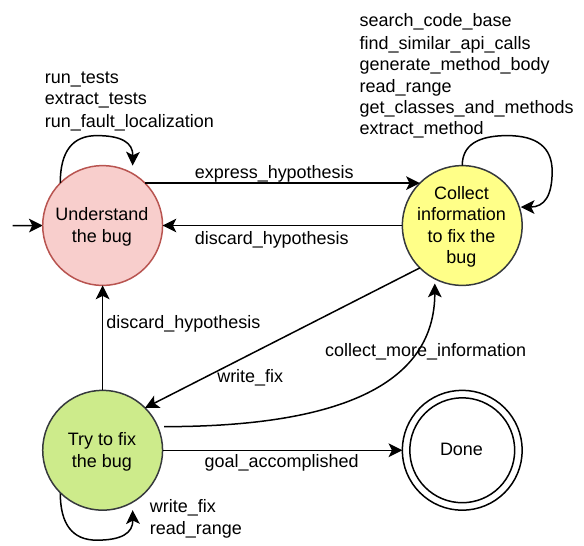}
	\caption{State machine to guide selection of tools.}
	\label{fig:statesdiagram}
\end{figure}

To guide the LLM agent toward using the available tools in an effective and meaningful way, we define a finite state machine that constrains which tools are available at a given point in time.
The motivation is that we observed the LLM agent to frequently get lost in aimless exploration in earlier experiments without such guidance.
Figure~\ref{fig:statesdiagram} shows the finite state machine, which we design to mimic the states a human developer would go through when fixing a bug.
Each state is associated with a set of tools available to the agent, which are described in Section~\ref{sec:tools}.
Importantly, the agent is free to transition between states at any point in time by using tools.
That is, despite providing guidance, the state machine does not enforce a strict order of tool invocations.

The state description section of the prompt informs the agent about its current state:
\begin{itemize}
	\item \emph{Understand the bug}:
	The agent starts in this state, where it can collect information related to the failing test cases and the bug's location. 
	Once the agent has an understanding of the bug, it formulates a hypothesis to describe the nature of the bug and the reason behind it.
	Throughout the repair process, the agent may refute earlier hypotheses and express new ones.
	After expressing a hypothesis, the agent will automatically switch to the next state.
		
	\item \emph{Collect information to fix the bug}:
	In this state the agent collects information that help suggest a fix for the bug expressed by the hypothesis, e.g., by searching for specific repair ingredients or by reading possibly relevant code.
	Once the agent has gathered enough information to attempt a fix, it can transition to the next state.
	
	\item \emph{Try to fix the bug}:
	In this state, the agent tries to fix the bug based on its current hypothesis and the collected information.
	Each fix attempt modifies the code base and is validated by executing the test cases.
	If necessarily, the agent can go back to one of the previous states to establish a new hypothesis or to gather additional information.
\end{itemize}
In addition to the three above states, \name{} has a final state, \emph{``Done''}, which the agent can reach by calling a specific command that indicates the success of repair.

\subsubsection{Available Tools}

This section of the prompt describes a set of tools that the agent can call at the current state.
Each tool has a name, a description, and a set of typed arguments (Section~\ref{sec:tools}).

\subsubsection{Gathered Information}

A key ability of the repair agent is to gather information about the bug and the code base, which serves as the basis for deciding which commands to invoke next.
To make this information available to the agent, we maintain a prompt section that lists the information gathered by the different tool invocations.
Intuitively, this section of the prompt serves as a memory for the agent, allowing it to recall information from previous cycles.
The gathered information is structured into different subsections, where each subsection contains the outputs produced by a specific tool.

\subsubsection{Specification of Output Format}

\begin{figure}
	\begin{lstlisting}[language=JSON]
interface Response {
  thoughts: string;
  command: {
    name: string;	
    args: Record<string, any>;
  };
}
	\end{lstlisting}
	\caption{JSON format of the response of the model.}
	\label{fig:output format}
\end{figure}

Given the dynamic prompt, the LLM agent provides one response per cycle.
To enable the middleware to parse the response, we specify the expected output format (Figure~\ref{fig:output format}).
The ``thoughts'' field provides a textual description of the agent's reasoning when deciding about the next command.
Asking the agent to express its thoughts increases the transparency and interpretability of the approach, provides a way to debug potential issues in the agent's decision-making process, and helps improve the reasoning abilities of LLMs~\cite{wei2022chain}.
The ``command'' field specifies the next command to be executed, consisting of the name of the tool to invoke and the set of arguments.

\begin{figure}
	\begin{lstlisting}[language=JSON]
{
  "thoughts": "With the hypothesis in mind, 
  it's time to gather more information to
  formulate potential fixes for the bug. I 
  should focus on understanding the context
  around the condition 
  'if (x < 0 && prev == '-')'
  to come up with effective solutions.",
  "command": {
    "name": "search_code_base",
    "args": {
  	  "key_words": 
		[ "addNumber", 
	 	  "CodeConsumer", 
 		  "if (x < 0 && prev == '-')" ]
}}}
		\end{lstlisting}
	\caption{Example of a response of the repair agent.}
	\label{fig:output example}
\end{figure}

For example, Figure~\ref{fig:output example} shows a response of the LLM agent.
The model expresses the need to collect more information to understand the bug and suggests a command that searches the code base with a list of keywords.

\subsubsection{Last Executed Command and Result}

This section of the prompt contains the last command (tool name and arguments) that was executed (if any) and the output it produced.
The rationale is to remind the agent of the last step it took, and to make it aware of any problems that occurred during the execution of the command. Furthermore, we remind the agent how many cycles have already been executed, and how many cycles are left.

\subsection{Tools for the Agent to Use}
\label{sec:tools}

A key novelty in our approach is to let an LLM agent autonomously decide which tools to invoke to fix a bug.
The tools we provide to the agent (Table~\ref{tab: commands}) are inspired by the tools that developers use in their IDEs.

\begin{table*}[]
	\centering
	\caption{Repair-related tools invoked by \name{}.}
	\label{tab: commands}
	\begin{tabular}{@{}p{2.8cm}p{13.3cm}@{}}
		\toprule
		Tool &
		Description \\
		\midrule
		\multicolumn{2}{@{}l}{\textbf{Read and extract code:}} \\
		\midrule
		\emph{read\_range} &
		Read a range of lines in a file. \\
		\emph{get\_classes\_and\_methods} &
		Get the names of all classes and methods in a file. \\
		\emph{extract\_method} &
		Given a method name, extract method implementations from a file.
		\\
		\emph{extract\_tests} &
		Given the failure report from JUnit or ANT, extract the code of failing test cases. \\
		\midrule
		\multicolumn{2}{@{}l}{\textbf{Search and generate code:}} \\
		\midrule
		\emph{search\_code\_base} &
		Scans all Java files within a project for a list of keywords.\\
		\emph{find\_similar\_api \_calls} &
		Given a code snippet that calls an API, search for similar API calls in the project. \\		
		\emph{generate\_method\_body} &
		Ask an LLM (GPT3.5 by default) to generate the body of a method based on code proceeding the method. \\
		\midrule
		\multicolumn{2}{@{}l}{\textbf{Testing and patching:}} \\
		\midrule
		\emph{run\_tests} &
		Run the test suite of a project. \\
		\emph{run\_fault\_localization} &
		Retrieve pre-existing localization information or run a fault localization tool.
		\\
		\emph{write\_fix} &
		Apply a patch to the code base and execute the test suite of the project. Changes are reverted automatically if tests fail. Moves the agent into the 'Try to fix the bug' state. \\		
		\midrule
		\multicolumn{2}{@{}l}{\textbf{Control:}} \\
		\midrule
		\emph{express\_hypothesis} &
		Express a hypothesis about the bug. Moves the agent into the 'Collect information to fix the bug' state. \\ 
		\emph{collect\_more\_information} &
		Move the agent back to the 'Collect information to fix the bug' state. \\ 
		\emph{discard\_hypothesis} &
		Discard the current hypothesis about the bug and move back to the 'Understand the bug' state. \\
		\emph{goal\_accomplished} &
		Declare that the goal has been accomplished and exiting the repair process. \\
		\bottomrule
	\end{tabular}
\end{table*}

\subsubsection{Reading and Extracting Code}

A prerequisite for fixing a bug is to read and understand relevant parts of the code base.
Instead of hard-coding the context provided to the LLM~\cite{Xia2023a,Kang2023a,Ye2024}, we let the agent decide which parts of the code to read, based on four tools.
The \emph{read\_range} tool allows the agent to extract a range of lines from a specific file, which is useful to obtain a focused view of a particular section of code.
To obtain an overview of the code structure, the \emph{get\_classes\_and\_methods} tool retrieves all class and method names within a given file.
By invoking the \emph{extract\_method} tool, the agent can retrieve the implementation(s) of methods that match a given method name within a given file.
Finally, we offer the \emph{extract\_tests} tool, which extracts the code of test cases that resulted in failure.
The tool is crucial to understand details of failing tests, such as input values and the expected output.

\subsubsection{Search and generate code}

Motivated by the fact that human developers commonly search for code~\cite{csur2023_code_search}, we present tools that allow the agent to search for specific code snippets.
These tools are useful for the agent to better understand the context of a bug and to gather repair ingredients, i.e., code fragments that could become part of a fix.
The \emph{search\_code\_base} tool enables the agent to locate instances of particular keywords within the entire code base.
For example, the agent can use this tool to find occurrences of variables, methods, and classes.
Given a set of keywords, the tool performs an approximate matching against all source code files in the project.
Specifically, the tool splits each keyword into subtokens based on camel case, underscores, and periods, and then searches for each subtoken in the code.
For example, searching for \code{quickSortArray} yields matches for \code{sortArray}, \code{quickSort}, \code{arrayQuickSort}, and other related variations.
The output of the tool is a nested dictionary, organized by file names, classes, and method names, that provides the keywords that match a method's content.
Another search tool, \emph{find\_similar\_api\_calls}, allows the agent to identify and extract usages of a method, which is useful to fix incorrect method calls.
Without such a tool, LLMs tend to hallucinate method calls that do not exist in the code base~\cite{arXiv2024_De-Hallucinator}.
Given a code snippet that contains a method call, the tool extracts the name of the called method, and then searches for calls to methods with the same name.
The agent can restrict the search to a specific file or search the entire code base.

In addition to searching for existing code, \name{} offers a tool that generates new code by invoking another LLM.
The tool is inspired by the success of LLM-based code completion tools, such as Copilot~\cite{Chen2021-short}, which human developers increasingly use when fixing bugs.
Given the code preceding a method and the signature of the method, the \emph{generate\_method\_body} tool asks an LLM to generate the body of the method.
The query to the code-generating LLM is independent of the dynamic prompt used by the overall \name{} approach, and may use a different model.
In our evaluation, we use the same LLM for both the repair agent and as the code-generating LLM of this tool.
The tool limits the given code context to 12k tokens and sets a limit of 4k tokens for the generated code.

\subsubsection{Testing and Patching}

The next category of tools is related to running tests and applying patches.
The \emph{run\_tests} tool allows the agent to execute the test suite of the project.
It produces a report that indicates whether the tests passed or failed.
In case of test failures, the tool cleans the output of the test runner, e.g., by removing entries of the stack trace that are outside of the current project.
The rationale is that LLMs have a limited prompt size and that irrelevant information may confuse the model.
The \emph{run\_fault\_localization} tool retrieves fault localization information, which is useful to understand which parts of the code are likely to contain the bug.
\name{} offers two variants of this tool: Either, it provides perfect fault localization information or it invokes an existing fault localization tool, such as GZoltar~\cite{campos2012gzoltar}, to calculate fault localization scores.
In case of perfect fault localization, the tool provides all the file(s) and line(s) that need to be edited to fix the bug.
As common in the field of program repair, we assume perfect fault localization as the default.

\begin{figure}
	\begin{lstlisting}[language=Json]
[
  {
    "file_path": "jfree/data/time/Week.java",
    "insertions": [
      {
      "line_number": 175,
      "new_lines": [
      "// ...new lines to insert...\n",
      "// ...more new lines...\n"]
      }
    ],
    "deletions": [179, 183],
    "modifications": [
      {
      "line_number": 179,
      "modified_line": "    if (dataset == null) {\n"
      }
    ]
  },
  {
    "file_path": "org/jfree/data/time/Day.java",
    "insertions": [],
    "deletions": [307],
    "modifications": []
  }
]
	\end{lstlisting}
	\caption{Example of patch given to the \emph{write\_fix} tool.}
	\label{fig:patch format}
\end{figure}

Once the agent has gathered sufficient information to fix the bug, it can apply a patch to the code base using the \emph{write\_fix} tool.
\name{} aims at repairing arbitrarily complex bugs, including multi-line and even multi-file bugs.
The \emph{write\_fix} tool expects a patch in a specific JSON format, which indicates the insertions, deletions, and modifications to be made in each file.
Figure~\ref{fig:patch format} shows an example of a patch in this format.
Given a patch, the tool applies the changes to the code base and runs the test suite.
If the tests fail, the \emph{write\_fix} reverts the changes, giving the agent a clean code base to try another fix.
Motivated by the observation that some fix attempts are almost correct, the \emph{write\_fix} tool requests the LLM to sample multiple variants of the suggested fix.
By default, \name{} samples 30 variants at max.
Given the generated variants, the approach removes duplicates and launches tests for every variant.

\subsubsection{Control}

The final set of tools do not directly correspond to a tool a human developer may use, but rather allow the agent to move between states (Figure~\ref{fig:statesdiagram}).
The \emph{express\_hypothesis} tool empowers the agent to articulate a hypothesis regarding the nature of the bug and to transition to the 'Collect information to fix the bug' state.
Inversely, the \emph{discard\_hypothesis} tool allows the agent to discard a hypothesis that is no longer viable, which leads back to the 'Understand the bug' state.
Together, the two commands enforce a structured approach to hypothesis formulation, aligning with work on scientific debugging~\cite{zeller2009programs,Kang2023a}.
In case the agent has tried multiple fixes without success, the \emph{collect\_more\_information} tool allows the agent to revert to the 'Collect information to fix the bug' state.
Finally, once the agent has found at least one fix that passes all tests, it can invoke the \emph{goal\_accomplished} tool, which terminates \name{}.

\subsection{Middleware}

The middleware component plays a crucial role in \name{}, orchestrating the communication between the LLM agent and the tools.
It performs the steps in Definition~\ref{def:cycle} as described in the following.

\subsubsection{Parsing and Refining LLM Output}

At the beginning of each cycle, the middleware queries the LLM with the current prompt.
Ideally, the response adheres perfectly to the expected format (Figure~\ref{fig:output format}).
In practice, the LLM may produce responses that deviate from the expected format, e.g., due to hallucinations or syntactic errors.
For example, the LLM may provide a ``path'' argument while the tool expects a ``file\_path'' argument.

\name{} tries to heuristically rectify such issues by mapping the output to the expected format in three steps.
First, it tries to map the tool mentioned in the response to one of the available tools.
Specifically, the approach checks if the predicted tool name $n_{\mathit{predicted}}$ is a substring of the name of any available tool $n_{\mathit{actual}}$, or vice versa, and if yes, considers $n_{\mathit{actual}}$ to be the desired tool.
In case the above matching fails, the approach checks if the Levenshtein distance between $n_{\mathit{predicted}}$ and any $n_{\mathit{actual}}$ is below a threshold (0.1 by default).
Second, the approach tries to map the argument names provided in the response to the tool's arguments, following the same logic as above.
Third, the approach handles invalid argument values by heuristically mapping or replacing them, e.g., by replacing a predicted file path with a valid one.
If the heuristics fail or produce multiple possible tool invocations, the middleware informs the LLM about the issue via the ``Last executed command and result'' prompt section and enters a new cycle.

In addition to rectifying minor mistakes in the response, the middleware also checks for repeated invocations of the same tool with the same arguments.
If the agent suggests the exact same command as in a previous cycle, the middleware informs the agent about the repetition and enters a new cycle.

\subsubsection{Calling the Tool}

Given a valid command from the LLM, the middleware calls the corresponding tool.
To prevent tool executions to interfere with the host environment or \name{} itself, the middleware executes the command in an isolated environment.

\subsubsection{Updating the Prompt}
Given the output of the tool, the middleware updates all dynamic sections of the prompt for the next cycle.
In particular, it updates the state description and the available tools, appends the tool's output to the gathered information, and replaces the section that shows the last executed command.

\section{Implementation}
We use Python 3.10 as our primary programming language. Docker is used to containerize and isolate command executions for enhanced reliability and reproducibility. \name{} is built on top of the AutoGPT framework and GPT-3.5-0125 from OpenAI. To parse and interact with Java code, we use ANTLR.

\section{Evaluation}
\label{sec:evaluation}

To evaluate our approach we aim to answer the following research questions:
\begin{enumerate}[label=RQ\arabic*,leftmargin=*]
\item How effective is \name{} at fixing real-world bugs?

\item What are the costs of the approach?

\item What is the influence and importance of the different components of \name{}?

\item How does the LLM agent use the available tools?
\end{enumerate}

\subsection{Experimental Setup}

\paragraph{Datasets}
We apply \name{} to bugs in the Defects4J dataset~\cite{Just2014}.
We use the entire Defects4J dataset, which consists of 835 real-world bugs from 17 Java projects, including 395 bugs from 6 projects in Defects4Jv1.2, as well as another 440 bugs and 11 projects added in Defects4Jv2.
Evaluating on the entire dataset allows us to assess the generalization capabilities of \name{} to different projects and bugs, without restricting the evaluation, e.g., based on the number of lines, hunks, or files that need to be fixed.

To assess the generalizability of our results and the potential influence of data leakage, we also evaluate \name{} on bugs from a newer dataset, GitBug-Java~\cite{gitbugjava}.
All bugs in this dataset were discovered and fixed in 2023, i.e., after the cut-off date of the GPT~3.5 version that we use in our evaluation (January 2022).
GitBug-Java contains 199 bugs from 55 projects.
Due to budget constraints, we randomly sample 100 of these bugs, sampling at least one and at most two bugs per project.
The random sample consists of 19 single-line bugs, 64 multi-line and 17 multi-file.

\paragraph{Baselines}
We compare with three existing repair techniques: ChatRepair~\cite{Xia2023a}, ITER~\cite{Ye2024}, and SelfAPR~\cite{Ye2022b}.
ChatRepair and ITER are two very recent approaches and have been shown to be the current state of the art.
All three baseline approaches follow an iterative approach that incorporates feedback from previous patch attempts.
Unlike \name{}, the baselines do not use an autonomous, LLM-based agent.
We compare against the baselines based on patches provided by the authors of the respective approaches.

\paragraph{Metrics of success}
Similar to past work, we report both the number of plausible and correct patches.
A fix is \emph{plausible} if it passes all test cases, but is not necessarily correct.
To determine whether a fix is correct, we automatically check whether it syntactically matches the developer-created fix.
If this is not the case, we manually determine whether the \name{}-generated fix is semantically consistent with the developer-created fix.
If and only if either of the two checks succeeds, we consider the fix to be \emph{correct}.

\subsection{RQ1: Effectiveness}

\subsubsection{Overall Results}

\begin{table}[]
	\centering
	\renewcommand{\arraystretch}{1.1}
	\caption{Results on Defects4J.}
	\label{tab:resultsd4j}
	\setlength{\tabcolsep}{1.5pt}
	\begin{tabular}{lrrr|rrr}
		\hline
		Project               & Bugs & Plausible & Correct & ChatRepair & ITER & SelfAPR \\ \hline
		Chart                 & 26   &    14      &   11        &       15                    & 10      &   7          \\
		Cli                   & 39   &    9      &     8        &        5                   &        6      &   8      \\
		Closure               & 174  &    27     &    27       & 37                          &      18      &   20        \\
		Codec                 & 18   &    10      &    9       &       8                    &        3      &   8     \\
		Collections           & 4    &    1      &     1      &        0                   &        0      &   1       \\
		Compress              & 47   &    10      &     10      &        2                   &        4      &   7       \\
		Csv                   & 16   &    6     &      6     &         3                  &         2      &   1      \\
		Gson                  & 18   &    3     &      3     &         3                  &         0      &   1      \\
		JacksonCore           & 26   &    5     &     5      &         3                  &         3     &   3      \\
		Jacksondatabind       & 112  &    18      &    11       &        9                   &        0      &   8       \\
		JacksonXml            & 6    &    1     &     1      &         1                  &         0     &   1      \\
		Jsoup                 & 93   &    18     &     18      &        14                   &        0     &   6       \\
		JxPath                & 22   &    0    &      0     &           0                &          0     &   1     \\
		Lang                  & 63   &    17      &    17       & 21                           &    0     &   10           \\
		Math                  & 106  &    29      &      29     & 32                           &    0     &   22           \\
		Mockito               & 38   &    6      &      6     & 6                          &       0     &   3        \\
		Time                  & 26   &     3     &      2     & 3                         &        2     &   3       \\ \hline
		Defects4Jv1.2         & 395  &   96       &     74      &          114         & 	57	&	64		\\
		Defects4Jv2         & 440  &    90      &     90     &         	48          &     --- 		&  46			\\
		Total          & 835  &    186      &     164      &        162           &		57			&	110	\\ \hline   
	\end{tabular}
\end{table}

\begin{table}[]
	\centering
	\renewcommand{\arraystretch}{1.2}
	\caption{Distribution of fixes by location type}
	\label{tab:bugstypes}
	\begin{tabular}{lr|rrr}
		\hline
		Bug type               & \name{} & ChatRepair & ITER & SelfAPR\\ \hline
		Single-line            &    115     &    133      &  36   & 83\\
		Multi-line* 			&    46     &    29      &  14   & 24\\
		Multi-file             &    3     &     0    &   4   & 3\\\hline
	\end{tabular}
\end{table}

Table~\ref{tab:resultsd4j} summarizes the effectiveness of \name{} in fixing the 835 bugs in Defects4J.
The approach generates plausible fixes for 186 bugs.
While not necessarily correct, plausible fixes pass all test cases and may still provide developers a hint about what should be changed.
\name{} generates correct fixes for 164 bugs, where 116 are exactly as fixed by the developers and 48 are semantically consistent with the developer-provided patches.
Being able to fix bugs from different projects shows that the approach can generalize to code bases of multiple domains.
Furthermore, \name{} creates fixes for bugs of different levels of complexity.
Specifically, as shown in Table~\ref{tab:bugstypes}, the approach fixes 115 single-line bugs, 46 multi-line (single-file) bugs, and 3 multi-file bugs.

\subsubsection{Comparison with Prior Work}

\begin{figure}[t]
	\centering
	\includegraphics[width=0.7\linewidth]{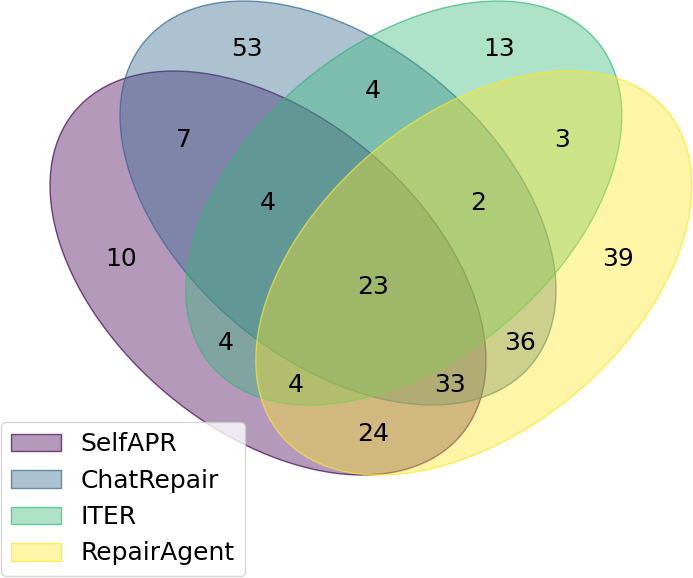}
	\caption{Intersection of the set fixes with related work. %
	}
	\label{fig:venndiag}
\end{figure}

The right-hand side of Table~\ref{tab:resultsd4j} compares \name{} with the baseline approaches ChatRepair, ITER, and SelfAPR.
Previous to this work, ChatRepair had established a new state of the art in APR by fixing 162 bugs in Defects4J.
\name{} achieves a comparable record by fixing a total of 164 bugs.
Our work particularly excels in Defects4Jv2, where \name{} fixes 90 bugs, while ChatRepair only fixes 48 bugs.
To further compare the sets of fixed bugs, Figure~\ref{fig:venndiag} shows the overlaps between different approaches.
As often observed in the field of APR, different approaches complement each other to some extent.
In particular, \name{} fixes 39 bugs that were not fixed by any of the three baselines.
Out of these 39 bugs, 18 are single-line, 20 are multi-line, and one is a multi-file bug.
Comparing the complexity of the bug fixes, as shown on the right-hand side of Table~\ref{tab:bugstypes}, \name{} is particularly more effective, compared to other tools, for bugs that require more than a single-line fix.
We attribute this result to the \name{}'s ability to autonomously retrieve suitable repair ingredients and the ability to edit an arbitrary number of lines and files.

\subsubsection{Examples}

\begin{figure}[t]
	\begin{lstlisting}[language=Java]
if (cfa != null) {
for (Node finallyNode : cfa.finallyMap.get(parent)) {
- cfa.createEdge(fromNode, Branch.UNCOND, finallyNode);
+ cfa.createEdge(fromNode, Branch.ON_EX, finallyNode);}}
	\end{lstlisting}
  \caption{Closure-14, bug fixed by \name{}.}
  \label{fig:example1}
\end{figure}

\begin{figure}[t]
\begin{lstlisting}[language=Java]
Separator sep = (Separator) elementPairs.get(0);
+ if (sep.iAfterParser == null && sep.iAfterPrinter == null) {
PeriodFormatter f = toFormatter(elementPairs.subList(2, size), notPrinter, notParser);
sep = sep.finish(f.getPrinter(), f.getParser());
return new PeriodFormatter(sep, sep);
+ }
\end{lstlisting}
  \caption{Time-27, bug fixed by \name{}.}
  \label{fig:example2}
\end{figure}

Figure~\ref{fig:example1} is a bug fixed exclusively by \name{}, where the agent uses the $find\_similar\_api\_calls$ tool to search for calls similar to \code{cfa.createEdge(fromNode, Branch.UNCOND, finallyNode);}. It returns a call from another file, which passes \code{Branch.ON\_EX} to the method call instead of \code{Branch.UNCOND}. This field name is then used as a repair ingredient by the agent.
In another example fixed only by \name{}, Figure~\ref{fig:example2}, \name{} benefitted from the $generate\_method\_body$ tool to generate a missing if-statement, which led to suggesting a correct fix afterwards.
These examples illustrate the clever and proper usage of available tools by the agent.
They also show these tools to be useful for finding repair ingredients that previous work fails to consider.

\subsubsection{Generalization and External Validity}

\begin{table}[t]
	\centering
	 \renewcommand{\arraystretch}{1.2}
	\caption{Results on GitBug-Java}
	\label{tab:gitbug}
	\begin{tabular}{lr|rrr}
		\hline
		Bug type               & Bugs & Plausible fixes & Correct fixes \\ \hline
		Single-line            &   19     &    11      &  9   \\
		Multi-line 			&    64     &    8      &  4   \\
		Multi-file             &    17     &     0    &   0   \\
		\hline
		Total & 100 & 19 & 13 \\ 
		\hline
	\end{tabular}
\end{table} 

To assess the generalization capabilities of \name{}, we evaluate the approach on GitBug-Java, with results shown in Table~\ref{tab:gitbug}.
Overall, \name{} finds 19 plausible fixes and 13 correct fixes.
The table shows that the approach is particularly effective for single-line bugs, where it correctly fixes 9 out of 19 bugs.
In contrast, the approach struggles with the 81 multi-line and multi-file bugs, where it finds only 4 correct fixes.
This result can at least partially be attributed to the fact that the GitBug-Java dataset contains more complex bugs than Defects4J.
The mean number of added and removed lines per ground truth bug fix are 2.9 and 9.3, respectively, for Defects4J, but 6.2 and 14.4 for GitBug-Java.
Likewise, the mean number of modified tokens is 381 for Defects4J, but 577 for GitBug-Java.
We conclude that \name{} generalizes well to new projects and bugs, and is not strongly affected by potential data leakage (e.g, Defects4J).

\subsection{RQ2: Costs of the Approach}

\begin{figure*}[t]
	\centering
	\captionsetup[subfigure]{aboveskip=-1pt,belowskip=1pt}
	\begin{subfigure}[b]{0.45\textwidth}
		\centering
		\includegraphics[width=\textwidth]{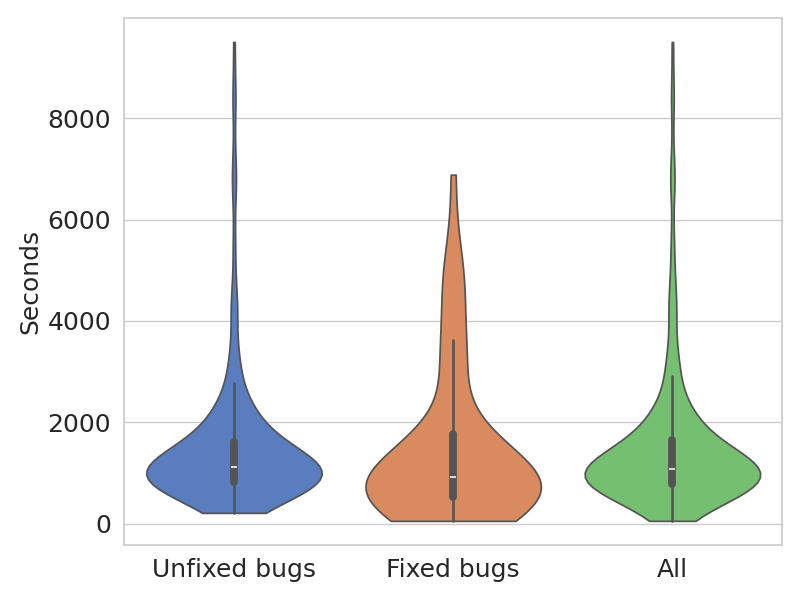}
		\caption{Time.}
		\label{fig:time}
	\end{subfigure}
	\begin{subfigure}[b]{0.45\textwidth}
		\centering
		\includegraphics[width=\textwidth]{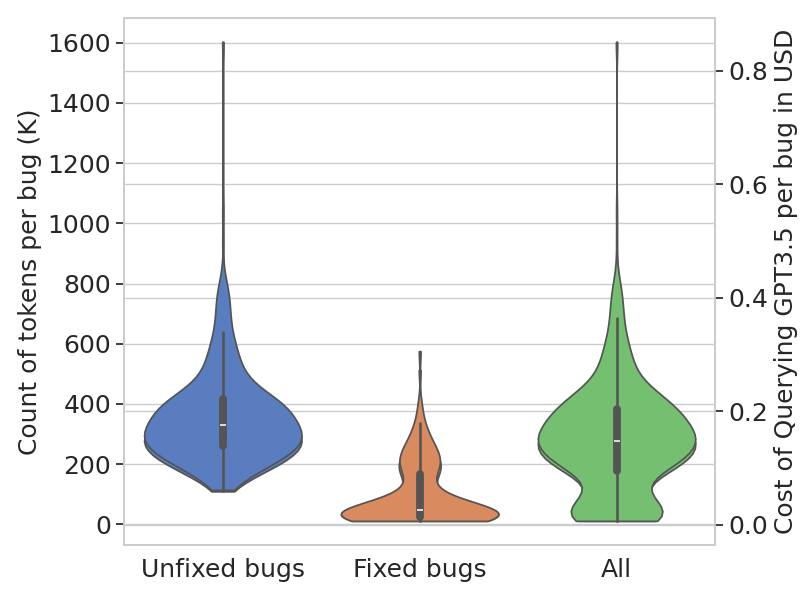}
		\caption{Tokens/money consumption.}
		\label{fig:cost}
	\end{subfigure}
	
	\caption{Distribution of cost metrics per bug (time, number of token, and monetary costs).}
	
	\label{fig:violincosts}
\end{figure*}

We measure three kinds of costs imposed by \name{}:
(i) Time taken to fix a bug.
(ii) The number of tokens consumed by queries to the LLM, which is relevant both for commercial models, such as the GPT-3.5 used here, and for self-hosted models, where the number of tokens determines the computational costs.
(iii) The monetary costs associated with the token consumption, based on OpenAI's pricing as of March 2024.

Our findings are summarized in Figure~\ref{fig:violincosts}.
The median time taken to address a bug is 920 seconds, with minimal variation between fixed and unfixed bugs.
Surprisingly, fixed bugs do not consistently exhibit lower repair times.
This is due to \name{}'s autonomous nature, where the repair process continues until the \emph{goal\_accomplished} command is invoked or the cycles budget is exhausted.
The figure shows several outliers where bug fixing attempt takes multiple hours.
\name{} spends 99\% of the total time in tool executions, mostly running tests.

Analyzing the costs imposed by the LLM, we find a median consumption of approximately 270,000 tokens, equating to around 14 cents (US dollars).
The number of tokens consumed by fixed bugs (21,000) is clearly lower than by unfixed bugs (315,000).
This difference is because the agent continues to extract additional information for not yet fixed bugs, saturating the prompt with operations, such as reading more lines of code.

\paragraph*{Comparison to prior work}
We compare the time and monetary costs of \name{} with other work based on what is reported in the respective papers.
The monetary costs of ChatRepair~\cite{Xia2023a} is reported as 42 cents per bug, based on the same model (GPT-3.5) as in our work.
Adjusting for the change in pricing between the two evaluations, the costs of ChatRepair would be 14 cents per bug under today's pricing, i.e., about the same cost as \name{}.
The monetary costs of ITER~\cite{Ye2024} and SelfAPR~\cite{Ye2022b} are not reported, as these approaches use self-trained models.
However, the authors of ITER report a median bug fixing time of 4.57 hours per bug, which is much higher than the median time of 920 seconds for \name{}.
While the comparison may be biased due to different hardware and software configurations, it suggests that \name{} is more efficient in terms of time costs.
We, mainly, attribute this difference to the number of patches that need to be validated (e.g., average of 117 patches generated by \name{} vs.\ 1000 patches generated by ITER).

\subsection{RQ3: Ablation Study}

\begin{table}[t]
	\centering
	\renewcommand{\arraystretch}{1.2}
	\caption{Different configurations of \name{}.}
	\label{tab:ablation}
	\setlength{\tabcolsep}{1.5pt}
	\begin{tabular}{lrrr|rrr}
		\hline
		Project               & Plausible & Correct & Cost(\$) & SL & ML & MF \\ \hline
		No search tools 		&	14	 &		11	&	28	&11	&	0&0 \\
		No state machine 		&	18	 &		14 &	31	&	9&	5& 0 \\
		Single-cycle memory		&	9	 &		6	&	8	&	5 &	1 & 0 \\
		Realistic localization 	&	16	 &		16	&	29	&	14&	 2&  0\\
		\hline
		\name{} (default) 		&	23	 &		21	&	16	& 16	&	5 & 0\\
		\hline   
	\end{tabular}
\end{table}

To better understand the impact of different components and configurations of \name{}, we perform the ablation studies summarized in Table~\ref{tab:ablation}.
Due to budget limitations, the ablations are done on a randomly selected set of 100 bugs of the entire Defects4J (same 100 for all configurations), out of which the full \name{} approach fixes 21 bugs.
We report the number of plausible and correct patches, costs in US dollars, and a break-down of correct fixes into single-line (SL), multi-line (ML), and multi-file (MF) bugs.

\paragraph*{Importance of search tools}
Without the search tools, \name{} fixes half of the bugs fixed by default.
The absence of search tools also causes the agent to read long sequences of code more frequently, which saturates the prompt quickly and doubles the costs.

\paragraph*{Importance of state machine}
Without guidance by the state machine (Figure~\ref{fig:statesdiagram}), the agent also fixes fewer bugs and has higher costs.
The main reason for the reduced effectiveness is that the agent does not follow a structured approach to fixing the bug.
For example, in many cases, the agent directly starts with suggesting a fix (often wrong) without collecting any information.

\paragraph*{Importance of long-term memory}
The third row of the table shows a variant of \name{} that keeps new information for a single cycle only, instead of accumulating all gathered information.
Again, the bug fixing effectiveness suffers significantly.
The reasons are that the agent repeats the same commands after a few cycles (e.g., to ask for the same information again), and it uses wrong file names and functions names.
Having a long-term memory helps the agent to keep useful information for future cycles, avoiding repeated queries.

\paragraph*{Impact of fault localization}
Finally, we evaluate \name{} with realistic fault localization, based on the spectrum-based GZoltar technique~\cite{campos2012gzoltar}. In total, \name{} fixes 16 bugs for 29 dollars which is a 25\% drop in fixing capability and a 81\% increase in costs. These results were achieved without giving \name{} more cycles which would have helped otherwise since the agent spends extra time on localizing the bug.

\subsection{RQ4: Usage of Tools by the Agent}

\begin{figure}
	\centering
	\includegraphics[width=1\linewidth]{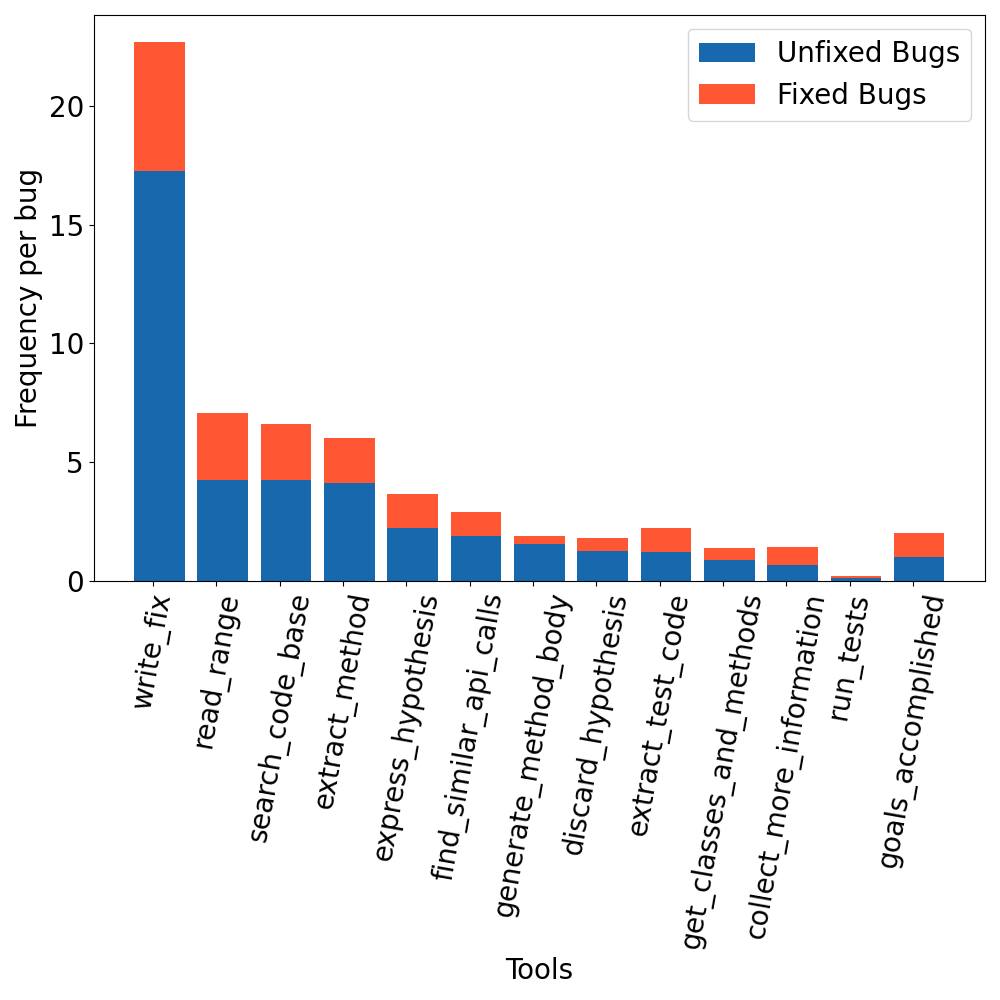}
	\caption{Frequency of tool invocations (average per bug).}
	\label{fig:stackedbarcommandfreq}
\end{figure}

This research question aims at better understanding the approach by analyzing how the agent uses the available tools.
On average, \name{} makes 35 tool invocations per bug, which also corresponds to the number of cycles.
Figure~\ref{fig:stackedbarcommandfreq} shows the frequency of tool invocations, where we distinguish between fixed (i.e., ``correct'') and unfixed (i.e., ``plausible'' only or completely unfixed) bugs.
The agent uses the full range of tools, with the most frequently called tool being \emph{write\_fix} (average of 6 calls for fixed bugs and 17 calls for unfixed bugs).
Around 7\% of \emph{write\_fix} invocations in unfixed bugs produce plausible patches, compared to 44\% in fixed bugs.
The least used tool is \emph{run\_tests}, which is used so infrequently because the initially provided information about the bug already provides information about any failing test cases and because the \emph{write\_fix} tool automatically invokes the test suite.

\section{Discussion}

\subsection{Qualitative Insights}
The following describes qualitative insights gained from inspecting \name{}'s logs.

\paragraph*{Understanding the bugs}
\name{}'s ability to actively retrieve information that helps understand a bug allowed to fix a new set of bugs without higher costs. Particularly, we observe four kinds of information to be useful:
(i) the code of failing test cases and the initial execution results, which we provide in the prompt of the first cycle;
(ii) code snippets retrieved by searching for similar code, e.g., using the \emph{find\_similar\_api\_calls} tool;
(iii) details about the code structure, such as the classes and methods in a file; and
(iv) feedback obtained by applying a fix, which triggers the test execution and reveals any test cases that still fail.

\paragraph*{Unfixed bugs and fix complexity}
As shown in Table~\ref{tab:bugstypes}, \name{} clearly outperforms prior work on multi-line bugs, but fails to fix some of the simpler, single-line bugs fixed by, e.g., ChatRepair~\cite{Xia2023a}.
We observe that the agent sometimes suggests complex fixes for bugs that only require simple modifications.
A possible remedy could be to initially limit the complexity of candidate fixes, nudging the agent toward trying simple fixes first.
For multi-line, multi-file bugs, we observe that \name{} often edits only a subset of the required locations.
Future work could explore human-in-the-loop approaches, where a partial fix found by an agent could give a developer a head start.

\subsection{Threats to Validity and Limitations}
\label{sec:threats}

While \name{} shows promising results, we acknowledge several potential threats to validity and inherent limitations:
\emph{(i) Data leakage:} GPT-3.5 may have seen parts of the Java projects we evaluate on during training. Our closest competitor, ChatRepair, also uses GPT-3.5, and thus faces the same risk.
Moreover, the experiment on GitBug-Java suggests that \name{} is effective also on bugs guaranteed to not be part of the training data.
\emph{(ii) Missing test cases:} Defects4J has at least one failing test case for each bug, which may not be the case for real-world usage scenarios.
It will be interesting to evaluate \name{} on bugs with no a-priori available error-revealing test cases in future work.
\emph{(iii) Fault localization:} Inaccurate or imprecise fault localization could lead to suboptimal repair suggestions or incorrect diagnoses.
\emph{(iv) Non-deterministic output of LLMs:} The inherently non-deterministic nature of LLMs may result in different outcomes between two consecutive runs of \name{}.
The large number of bugs we evaluate on mitigates this risk.
Moreover, the logs of interactions with the LLM are available for further analysis.

\section{Related Work}
\label{sec: rw}

\paragraph{Non-learning-based program repair}
Automated program repair~\cite{cacm2019-program-repair} has received significant attention.
Some approaches address it as a search problem based on manually designed code mutation rules and fix patterns~\cite{LeGoues2012,Le2016,Liu2019a}.
Alternatively, transformation rules can be derived (semi-)automatically from human-written patches~\cite{Kim2013,oopsla2019,Bavishi2019}.
Other approaches use symbolic constraints to derive fixes~\cite{Nguyen2013b,Ke2015,xuan2016nopol,mechtaev2016angelix}, integrate repair into a static analysis that identifies bugs~\cite{Tonder2018,Liu2023,Jain2023}, or replace buggy code with similar code from the same project~\cite{Yang2022}.
APR has been successfully deployed in industrial contexts~\cite{oopsla2019,Marginean2019}.
Beyond functional bugs, several techniques target other kinds of problems, such as syntax errors~\cite{Wang2018,Gupta2019a,Sakkas2022}, performance bugs~\cite{emse2017}, vulnerabilities~\cite{Harer2018}, type errors~\cite{icse2024-PyTy}, common issues in deep learning code~\cite{Zhang2021}, and build errors~\cite{Tarlow2020}.

\paragraph{Learning-based program repair}
While early work uses machine learning to rank and select candidate fixes~\cite{Long2016}, more recent work uses machine learning to generate fixes.
Approaches include neural machine translation models that map buggy code into fixed code~\cite{Gupta2017,Tufano2019,Lutellier2020,Chen2021d}, models that predict tree transformations~\cite{Li2020a,Zhu2021}, neural architectures for specific kinds of bugs~\cite{Vasic2019}, and repair-specific training regimes~\cite{Ye2022a,Ye2022b}.
We refer to a recent survey for a more comprehensive discussion~\cite{zhang2023survey}.
Unlike the above work, \name{} and the work discussed below use a general-purpose LLM, instead of training a task-specific model.

LLMs have motivated researchers to apply them to program repair, e.g., in studies that explore prompts~\cite{Jiang2023,10172803} and in a technique that prompts the model with error messages~\cite{Joshi2023}.
These approaches perform a one-time interaction with the model, where the model receives a prompt with code and produces a fix.
The most recent techniques introduce iterative approaches, which query the LLM repeatedly based on feedback obtained from previous fix attempts~\cite{Xia2023a,Kang2023a,Ye2024,hidvegi2024cigar}.
\name{} also queries the model multiple times, but fundamentally differs by pursuing an agent-based approach.
Section~\ref{sec:evaluation} empirically compares \name{} to the most closely related iterative approaches~\cite{Xia2023a,Ye2024}.

\paragraph{LLMs for code generation and code editing}
Beyond program repair, LLMs have been applied to a variety of other code generation and code editing tasks, including
code completion~\cite{Chen2021-short,Shrivastava2023},
fuzzing~\cite{icse2024-Fuzz4All},
generating and improving unit tests~\cite{Lemieux2023,DBLP:journals/tse/SchaferNET24,Ryan2024,Alshahwan2024,Kang2023,Feng2024},
multi-step code editing~\cite{Bairi2023}.
Unlike our work, none of these approaches uses an agent-based approach.

\paragraph{LLM-based agents}
The idea to let LLM agents autonomously plan and perform complex tasks is relatively new and has been applied to tasks outside of software engineering~\cite{Wang2023}.
To the best of our knowledge, our work is the first to apply an LLM-based agent to program repair or any other code generation problem in software engineering.
Copra is an agent-based approach for formal theorem proving~\cite{thakur2024incontextlearningagentformal}. 
After an initial version of this paper was made publicly available, other LLM-based agents for software engineering tasks have been proposed~\cite{Zhang2024a,Yang2024a}, showing the potential of this kind of approach.
\name{} is inspired by prior work~\cite{Mialon2023} on augmenting LLMs with tools invoked via APIs~\cite{Schick2023,Patil2023} and with the ability to generate and execute code~\cite{Gao2022}.
Our key contribution in applying these ideas to a software engineering task is to define tools that are useful for program repair and a prompt format that allows the LLM to interact with these tools.

\section{Conclusion}
This paper presents a pioneering technique for bug repair based on an autonomous agent powered by Large Language Models (LLMs). Through extensive experimentation, we validate the effectiveness and potential of our approach. Further exploration and refinement of autonomous agent-based techniques will help generalize to more difficult and diverse types of bugs if equipped with the right tools.

\bibliographystyle{IEEEtran}
\bibliography{mainrefs.bib,referencesMP.bib}
\end{document}